\documentclass[preprint,prb,aps,amsmath,amssymb,groupedaddress,showpacs,letterpaper]{revtex4}

\usepackage {graphicx}

\def \oli    {LiMPO$_4$}
\def \po     {MPO$_4$}
\def \olix   {Li$_x$MPO$_4$}

\begin{document}
\title{
Towards more accurate First Principles prediction of redox potentials in transition-metal compounds with
LDA+U}

\author{F. Zhou}
\affiliation{Department of Physics,  Massachusetts Institute of Technology, Cambridge, MA 02139}
\author{M. Cococcioni, C. A. Marianetti, D. Morgan, G. Ceder}
\affiliation{Department of Material Science and Engineering, Massachusetts Institute of Technology, Cambridge, MA 02139}

\date{\today}

\begin{abstract}
First-principles calculations within the Local Density Approximation (LDA) or Generalized Gradient
Approximation (GGA), though very successful, are known to underestimate redox potentials,
such as those at which lithium intercalates in transition
metal compounds.
We argue that this inaccuracy is related to the lack of cancellation of electron self-interaction errors in LDA/GGA and can be improved
by using the DFT+$U$ method with a self-consistent evaluation of the $U$ parameter. We show that, using this approach,
the experimental  lithium intercalation voltages of a number of transition metal compounds,
including the olivine
Li$_{x}$MPO$_{4}$ (M=Mn, Fe Co, Ni), layered Li$_{x}$MO$_{2}$ ($x=$Co, Ni)
and spinel-like Li$_{x}$M$_{2}$O$_{4}$ (M=Mn, Co),
can be reproduced accurately.
\end{abstract}

%
%
\maketitle

\section{Introduction}

Redox processes are relevant to many technological applications, including corrosion,
fuel cells and rechargeable Li batteries,
and the ability to study these processes from first principles is therefore crucial.
The key to a redox reaction is the transfer of electrons from one species to another. When the
redox electron is transferred between very distinct environments (e.g. metallic to ionic)
the standard Local Density Approximation (LDA) and Generalized Gradient
Approximation (GGA) lead to considerable errors in the calculated redox energies.
We show in this paper that treating self-interaction with the DFT+$U$ \cite{Anisimov1991, Anisimov1993, Liechtenstein1995}
method gives considerably better agreement with experiment and thereby provides
a tool to accurately predict redox potentials.

In particular, we focus on the study of Li insertion
in transition metal compounds using GGA and GGA+$U$. Transition metal (TM) compounds
have attracted intense research  as cathode
materials for rechargeable Li batteries due to their ability to simultaneously absorb Li$^+$ ions and electrons.
In the discharge cycle of a rechargeable  battery  Li is oxidized on the anode side and inserted as Li$^{+}$ + e$^{-}$
in the TM compound that comprises the cathode.
The energy of this reaction determines the oxidization/reduction potential at which
the battery operates. It is the high redox potential of Li cells that makes them so desirable in applications
where high energy density is required.

First principles calculations have been used extensively to predict important properties of Li-insertion materials such as
the average potential \cite{Aydinol1997, Wolverton-voltage1998, Launay-DFT-Li2003, Tang-Fe_phosphate2003, Hwang2003, Rocquefelte-LiV2O5-2003, Kganyago-LiC6-2003, Prasad-MnO2003, Koyama2003, Morgan2002, Doublet2002}
and potential  profile  \cite{Anton-phase1998,Courtney-profile1998} for Li insertion,
phase stability \cite{Anton-phaselicoo2-1998,Elena-LiNiO2-2002, Reed-spinel2001} and Li diffusion \cite{Anton-diffusion2000}.
While this has led to considerable success in predicting the trends of Li insertion voltages \cite{Aydinol1997}
and even new phases \cite{Anton-phase1998}, it has been noted that LDA or GGA can give relatively large errors
for the average Li insertion potential \cite{Aydinol1997, Dane-NASICON2003}.
For example, Table \ref{tab:error} compares the experimental voltage for different structures with the
one calculated in the GGA approximation and with computational details discussed in section \ref{calc}.
\begin{table}[htbp]
\begin{tabular}{|c|c|c|c|}
  \hline     &  LiNiO$_2$/NiO$_2$ &LiMn$_2$O$_4$/Mn$_2$O$_4$   &LiFePO$_4$/FePO$_4$  \\
  \hline GGA & 3.19               &  3.18                      &  2.97               \\
  \hline exp.& 3.85\cite{Delmas-LA_Ni1999}        &  4.15 \cite{Ohzuku-LA_Co1994}              &  3.5 \cite{Yamada-Fe_phosphate2001}       \\
  \hline
\end{tabular}
\caption{Calculated and experimental redox couple voltage in Volt.}
\label{tab:error}
\end{table}
The Li insertion potential is consistently underpredicted by as much as 0.5 to 1.0V. Similar results have been obtained with LDA \cite{Aydinol1997}.

Recently, we have shown that electron correlation plays an important role
in predicting the phase diagram of the Li$_{x}$FePO$_{4}$ system \cite{Zhou-olivine2004},
for which LDA and GGA qualitatively fail.
In this work we demonstrate that the DFT+$U$ method also corrects the voltage error from LDA and GGA. In our
approach, $U$ is calculated self-consistently \cite{Cococioni-thesis2002}, thereby making this a
``first-principles'' approach to predict redox potentials with no adjustable parameters.

We first present some background information on the specific Li insertions  materials  investigated
and how the electrochemical
reactions take place in a rechargeable lithium battery.
We also discuss the details of the DFT+$U$ method and the self-consistent calculation of $U$.
In Sec. \ref{calc} we show the results of our approach, highlighting the improvement over GGA and
the good agreement with experiment.

\section{Materials and Methodology}
\subsection{Materials and Crystal structures}

As a representative set of Li-insertion compounds, we
have selected several materials representing different environments for Li and TM ions, which
are well characterized experimentally.

\begin{figure}[htbp]
\includegraphics[width=0.7 \linewidth]{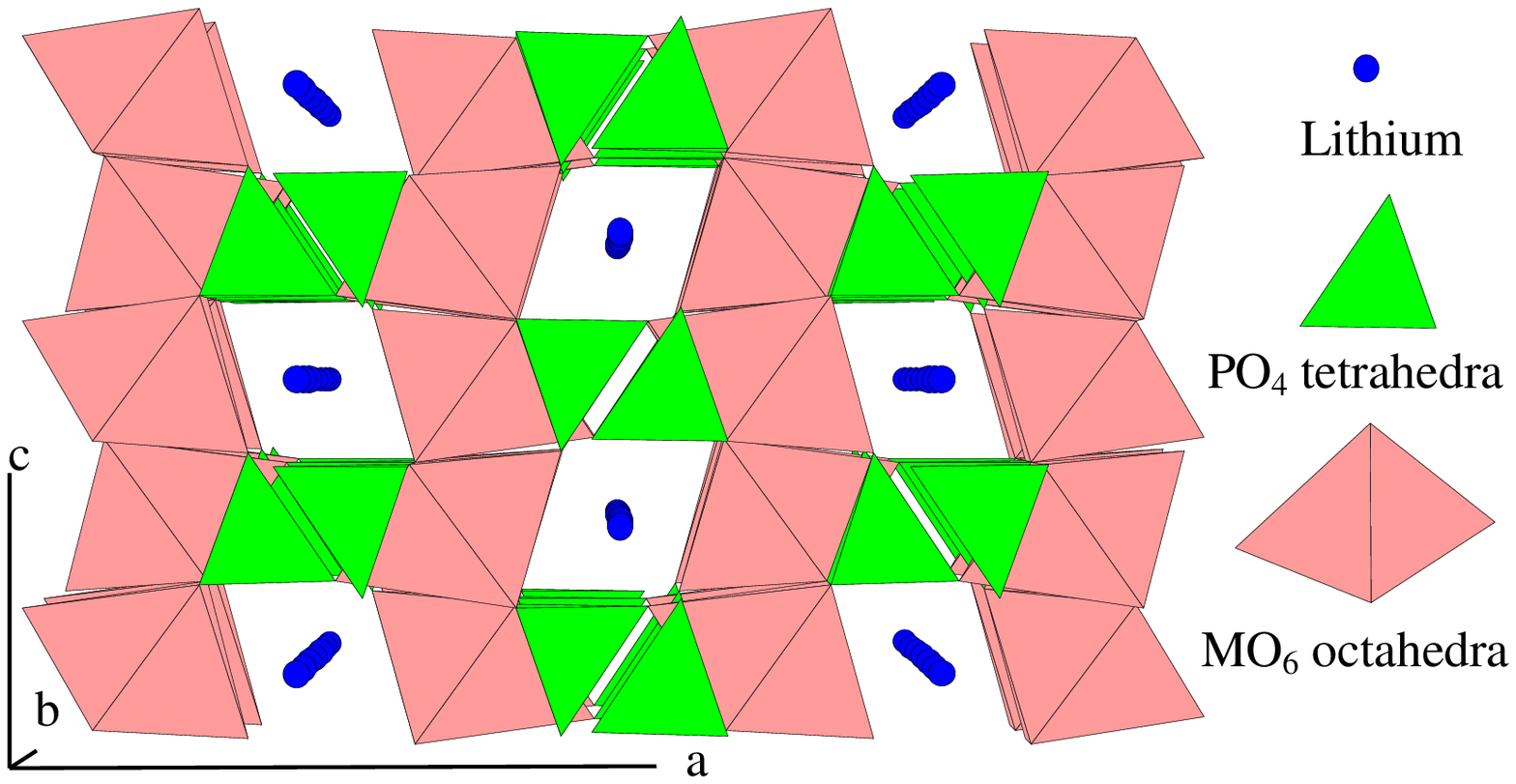}
\caption{The olivine structure with cation polyhedra.
\label{fig:olivine-structure}}
\end{figure}

The family \oli\ of olivine structures (M=Mn, Fe, Co and Ni) is a promising candidate for rechargeable Li-battery
electrodes in large applications such as electric and hybrid vehicles \cite{Goodenough1997}.
Olivine-type \oli\ and the de-lithiated structure \po, have an orthorhombic
unit cell with four Formula Units (FU) and space group $Pnma$ (see Fig.\ \ref{fig:olivine-structure}).
The olivine structure can be thought of as a distorted hexagonal close-packing of
oxygen anions, with three types of cations occupying the interstitial sites:
1) Transition metals M in corner-sharing MO$_{6}$ octahedra which are nearly coplanar to form a distorted
2-d square lattice perpendicular to the {\bf a} axis,
2) Lithium ions in edge-sharing LiO$_{6}$ octahedra aligned in parallel chains along the {\bf b} axis,
and 3) P ions in tetrahedral PO$_{4}$ groups connecting neighboring planes or arrays.
It is believed that the PO$_{4}$   groups 
hybridize less with the TM than an oxygen anion does in simple
close-packed oxides, and hence leads to more localized 3$d$ states
on the TM than in an oxide.

\begin{figure}[htbp]
\includegraphics[width=0.7 \linewidth]{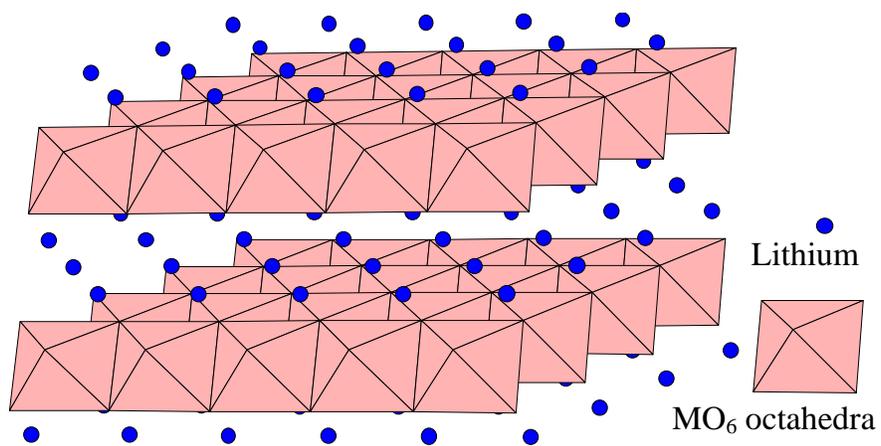}
\caption{The layered structure with MO$_{6}$ octahedra and lithium atoms.
\label{fig:layered-structure}}
\end{figure}

\begin{figure}[htbp]
\includegraphics[width=0.58 \linewidth]{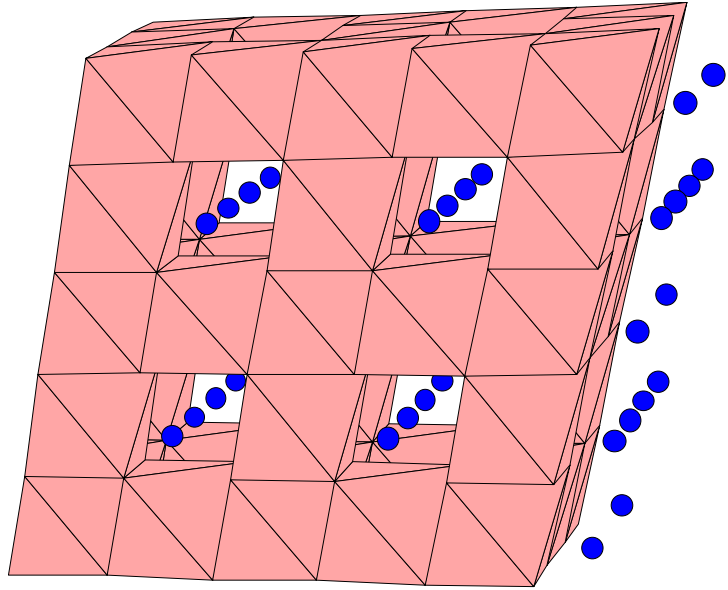}
\caption{The spinel-like structure when fully lithiated ($x=2$, Li atoms
taking octahedral positions) with MO$_{6}$ octahedra and lithium atoms.
\label{fig:sp-structure}}
\end{figure}

The layered LiMO$_{2}$ and spinel-like Li$_{x}$M$_{2}$O$_{4}$  are more traditional cathode
materials that have been thoroughly studied experimentally \cite{Thackeray-review1995}
and theoretically \cite{ Aydinol1997,Wolverton-voltage1998,Anton-diffusion2000,Reed-spinel2001}.
They are both ordered rock salts (see Figs. \ref{fig:layered-structure}, \ref{fig:sp-structure}).
The layered structure can be envisioned as two interpenetrating fcc lattices, one consisting
of oxygen, and the other consisting of alternating (111) planes of Li and TM.
In the R$\bar{3}$m space group the Li and the metal ions remain fixed in the ideal rock
salt positions, but the whole (111) oxygen planes can relax in the [111] direction.
The spinel-like structure Li$_{x}$M$_{2}$O$_{4}$ is so named because at $x=1$ it has the
same structure as the spinel mineral MgAl$_{2}$O$_{4}$. We shall refer to it as spinel
even when $x=2$. It can be envisioned as a fcc oxygen sublattice, with TM
in half of the octahedral oxygen interstices,
and  lithium either in part of the tetrahedral sites at $x=1$
or in the  octahedral sites not occupied by the TM ions at $x=2$ \cite{Thackeray-review1995}.

\subsection{Relation between insertion voltage and total energies}

When Li is inserted into a TM-oxide, its  charge is compensated by an electron absorbed from the external circuit.
The insertion reaction is symbolized by the following equation:
\begin{equation}
\Delta x \mathrm{Li +Li}_x \mathrm{MO_y} \Longleftrightarrow \mathrm{Li}_{x+ \Delta x}\mathrm{MO_y},
\label{eq:reaction}
\end{equation}
 where MO$_{y}$ is the TM compound host material.
Using thermodynamical arguments, it is possible to relate the
voltage $V$ of the cell to the lithium chemical potential ($\mu_{Li}$)
on both sides of Eq. \ref{eq:reaction} in the cathode \cite{McKinnon:book}:

%
\begin{equation}
V(x)= - \frac{\mu_{Li(x)}^{cathode}-\mu_{Li}^{anode}}{F}.
\label{eqn:voltage-x}
\end{equation}
%
F is the Faraday constant, and $\mu_{Li}^{anode}$ is the chemical potential in the anode, or more generally,
the chemical potential of the Li source.

The average voltage $\langle V \rangle$ for Li insertion between two composition limits, Li$_{x_1}$MO$_{y}$
and  Li$_{x_2}$MO$_{y}$,  can be found by integrating Eqn.\ \ref{eqn:voltage-x}
 (usually between $x=0$ and 1),
and is determined by the free energy of the compounds at the composition limits \cite{Aydinol1997}.
Neglecting the entropic and $P\Delta V$ contributions \cite{Aydinol1997}, $\langle V \rangle$
can simply  be determined by computing the total energy of Li$_{x_2}$MO$_y$,
Li$_{x_1}$MO$_{y}$  and Li:
%
\begin{equation}
\langle V \rangle = \frac{-\left[ E({Li_{x_2}MO_y}) - E({Li_{x_1}MO_y})  -(x_2-x_1)E({Li\ metal})    \right]}{ (x_2 -x_1) F}.
\label{eq:voltage-energy}
\end{equation}
Typically $x_1=0$ and $x_2=1$ are taken as composition limits, as in these cases no
Li-vacancy disorder occurs.

Experimentally, the voltage vs. lithium composition curve $V(x)$ can be conveniently
measured for both the charging and the discharging processes. The corresponding
curves differ in general because of the overcharge potential present in
the circuit. We obtain the experimental average open circuit voltage values by numerically averaging
the charge and discharge curves published in Refs. \onlinecite{Li-Mn_phosphate2002, Yamada-Fe_phosphate2001, Amine-Co_phosphate2000,
Ohzuku-LA_Co1994, Delmas-LA_Ni1999, Choi-SP_Co2002}
over the appropriate composition range.

\subsection{The DFT+U method}

The DFT+U method, developed in the 1990's \cite{Anisimov1991, Anisimov1993, Liechtenstein1995}, is now a well-established model
to deal with electron correlation in TM and rare earth compounds.
The method combines the high efficiency of LDA/GGA, and an explicit treatment of correlation
with a Hubbard like model for a subset of states in the system.
To investigate whether the underestimation of the lithium intercalation voltage in LDA/GGA could be related to Coulombic on-site effects
we carried out rotationally invariant DFT+$U$ \cite{Liechtenstein1995} calculations.
The essence of the method can be summarized by the expression for the total energy
\begin{equation}
\label{eq:ldapu}
E_{{\rm LDA+}U}[\rho, \hat{n}]= E_{\rm LDA}[\rho] + E_{\rm Hub}[\hat{n}] - E_{\rm dc}[\hat{n}]
        \equiv  E_{\rm LDA}[\rho] +E_{U}[\hat{n}]
\end{equation}
where $\rho$ denotes the charge density and $\hat{n}$ is the TM on-site $3d$ occupation matrix.
For these states the Hubbard interaction term $E_{\rm Hub}$ replaces
the LDA energy contribution $E_{\rm dc}$. The  $U$ correction term $E_{U}\equiv E_{\rm Hub} - E_{\rm dc}$
is defined by Eq. \ref{eq:ldapu}. Although  $E_{\rm dc}$ is not uniquely defined, we have chosen the spherically
averaged version \cite{dudarev} due to the considerations discussed in Ref. \onlinecite{Zhou-olivine2004}:
\begin{eqnarray}
E_{\rm dc}(\hat{n})  &=& \frac {U-J} 2  {\rm Tr}\hat{n}( {\rm Tr}\hat{n}-1)= \frac {U_{\rm eff}} 2  {\rm Tr}\hat{n}( {\rm Tr}\hat{n}-1), \\  \label{eq:dc}
E_{U}(\hat n)&=& \frac {U-J} 2 {\rm Tr}\left(\hat{n} (1-\hat{n}) \right)= \frac {U_{\rm eff}} 2 {\rm Tr}\left(\hat{n} (1-\hat{n}) \right), \label{eq:E_u}
\end{eqnarray}
where we have defined the effective interaction parameter $U_{\rm eff}=U-J$, or simply $U$ afterwards.
The calculated energies are insensitive to the $J$ parameter at fixed $U_{\rm eff}$ \cite{Zhou-olivine2004} and we include
it in $U_{\rm eff}$.


\subsection{Self-consistent Calculation of effective $U$}

We determine the $U$ parameter using the method presented in Ref.\ \onlinecite{Cococioni-thesis2002} which we briefly outline below.
This method is based on calculating the response in the occupation of TM states to a small perturbation of their
local potential.

We start from an LDA/GGA ($U=0$) calculation as the reference point. Then a small perturbation
$$dV=\alpha P_d^i,\ \ \   P_d^i=\sum_{m=-2}^{2} | m^i \rangle \langle m^i |  $$
 in the local
d-orbital potential is exerted on metal site $i$, where $P_d^i$ represents the projector on the $d$ states manifold of ion $i$,
and $\alpha$ is the amplitude of the potential shift applied to the $d$ levels. This induces a change in the occupation number of
ion $i$ as well as other ions. Thus we can calculate directly the response matrices,
\begin{equation}
\chi_{ji} = \frac{dn_d^j}{d \alpha_i},\ \ \   \chi_{0ji} = \frac{dn_{0d}^{j}}{d \alpha_{i}}   , \label{eq:response}
\end{equation}
which measure the variation of the $d$-manifold charge density $n_d^j$, on ion $j$,
produced by a potential shift at ion $i$.
The subscript ``0'' denotes the bare response, calculated without self-consistency (the
 Kohn-Sham potential apart from $dV$ is frozen at the value obtained in LDA/GGA before the perturbation), and corresponds
to the response from an independant electron system, while $\chi_{ji}$ is
the screened response (charge density and potential relaxed to reach self-consistency). The effective interaction parameter $U$ is then obtained as
\begin{equation}
U= (\chi_0^{-1} - \chi^{-1})_{ii}. \label{eq:diff-response}
\end{equation}
This is a well-known result in linear response theory,
in which the effective electron-electron interaction kernel is given as a difference among the interacting density response and the non-interacting
one \cite{Adler-response1962}. Since DFT is used, a finite contribution from the exchange-correlation potential is also
included in the effective $U$.
As we use the integrated quantity $n_d^i$ to probe the responses, the calculated effective interaction is averaged over
the ion in the same spirit as DFT+$U$.
The matrix in Eq. \ref{eq:diff-response}, whose diagonal term defines the on-site Hubbard $U$, also contains non-diagonal
terms corresponding to inter-site effective interactions in LDA/GGA. These are not used in the DFT+$U$ model.
This method to compute $U$ is contains full account of the screening to the
external perturbation operated by the electron-electron interactions.
In fact the perburtation is applied in larger and larger supercells until convergence of calculated $U$ is reached.
We also notice that the calculation of $U$ is based on the use of the same occupancy matrices
entering the DFT+$U$ functional, guaranteeing full consistency with the calculation we perform \cite{Cococioni-thesis2002}.

\section{Details of the calculation and Results \label{calc}}

Total energy calculations are performed for Ni, Mn, Co and Fe in the olivines, layered and spinel structures
whenever experimental information on the voltage is available.
For each system the total energy of the lithiated and delithiated
state is calculated with GGA and GGA+$U$, with the
projector-augmented wave (PAW) method \cite{paw, paw-vasp} as
implemented in the Vienna Ab-initio Simulation Package
\cite{vasp1}. The use of GGA over LDA has previously been shown to
be essential for correctly reproducing magnetic interactions and
possible Jahn-Teller distortions \cite{Mishra-JT-GGA1999}. An
energy cut-off of 500 eV and appropriate $k$-point mesh were
chosen so that the total ground state energy is converged to
within 3meV per FU. All atoms and cell parameters of each
structure are fully relaxed. Jahn-Teller distortions are allowed
where the transition metal
 ions are Jahn-Teller active (Mn$^{3+}$ and Ni$^{3+}$ in our case)
by explicitly breaking the symmetry of the unit cell.
Our relaxed cells of layered LiNiO$_{2}$ and spinel Li$_2$Mn$_2$O$_4$ agree well with
the calculations in Ref. \onlinecite{Marianetti-JT-2001} on Jahn-Teller distorted systems using GGA.
All calculations are performed with spin-polarization.
As discussed later, the total energy of a given structure depends critically on the magnetic
state of the metal ions, and high-spin states are favored by the DFT+$U$ scheme we used.
The ordering of the spin on the ions in different magnetic structures (i.e. ferromagnetic,
antiferromagnetic or more complicated ordering) results in difference in the total energy
of the order 10--60 meV per formula unit. From the total energies, the average lithiation potential can be calculated through Eqn.\ \ref{eq:voltage-energy}.

Table \ref{tab:U} shows the self-consistently calculated effective $U$
values for Mn, Fe, Co and Ni in different
valence states and structures. For each structure, $U$ is
calculated for the low and high valence states  respectively
in a fully lithiated and de-lithiated structure.
In all cases, except Ni$^{3+}$/Ni$^{4+}$ in the layered structure,
a higher valence state leads to a higher $U$. For the three cases
(Mn$^{3+}$/Co$^{3+}$/Ni$^{3+}$) for which we have a $U$ in a
close-packed (layered or spinel) oxides and in an olivine
phosphate structure, $U$ is higher for the olivine structure.
This may be related to the fact that the TM-octahedra in the
olivine are only corner sharing in two directions but separated
from each other by phosphate groups in the third direction,
leading to very narrow bandwidth and well localized TM-$d$ states.
For comparison we also list the $U$ values calculated in Ref.
\onlinecite{Pickett-calcU1998} for TM monoxides MO (M = Mn, Fe, Co and
Ni) in non spin-polarized state. Good agreement with \oli\ is
found except for Fe$^{2+}$. 
We note that in Ref \onlinecite{Cococioni-thesis2002} the $U$ value of 4.3 eV for FeO was
obtained with the same linear response approach, in good agreement with Ref \onlinecite{Pickett-calcU1998}.
So the difference between our results and those in Ref \onlinecite{Pickett-calcU1998} 
could be mainly due to different crystal environment.

\begin{figure}[htbp]
\includegraphics[width=0.7 \linewidth]{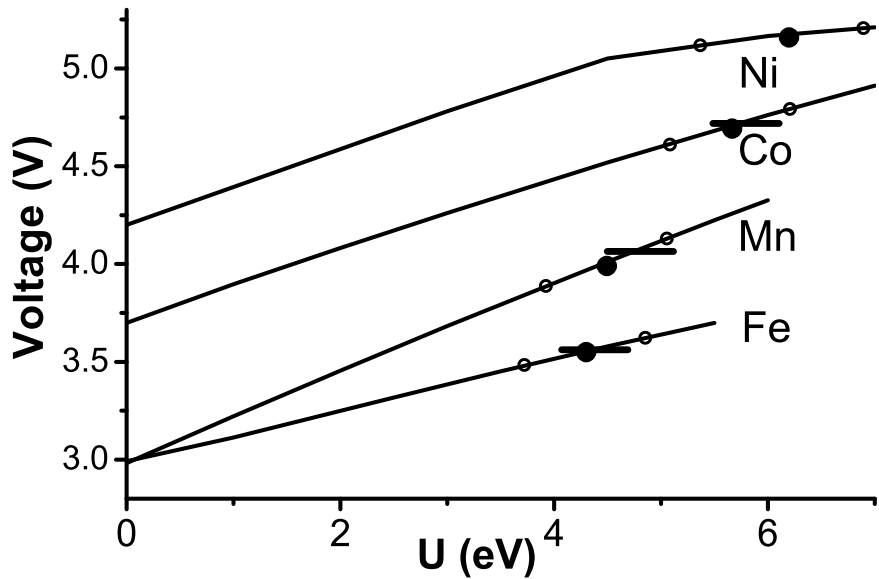}
\caption{
Voltage as a function of $U$ for the \oli\ materials in the olivine structure. The short horizontal
lines on the curves indicate the experimental voltage of the
each material (no experimental information is available for LiNiPO$_{4}$).
The two small open circles on a curve represent the voltage for $U$ calculated
in the oxidized (delithiated) or reduced (lithiated) states. The big solid circle
represents the voltage at the average of the two $U$ values.
}
\label{fig:voltage-u-ol}
\end{figure}
\begin{figure}[htbp]
\includegraphics[width=0.7 \linewidth]{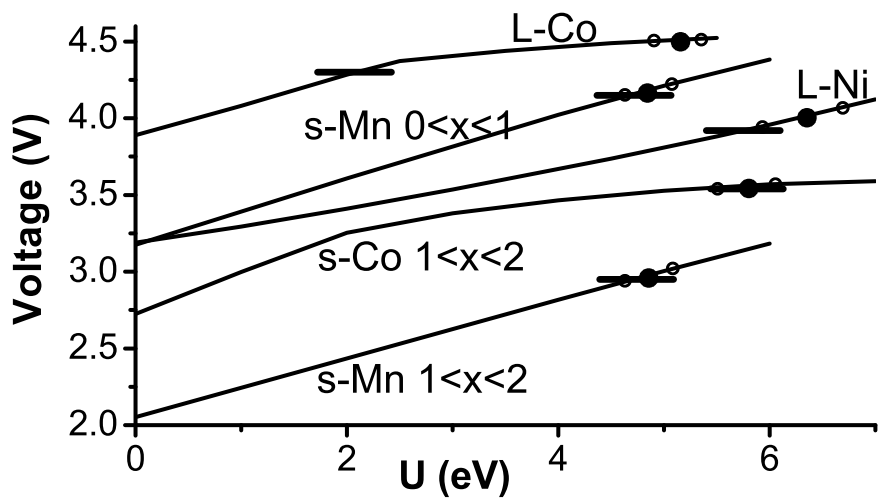}
\caption{
Voltage as a function of $U$ for the layered and spinel structures.
Legend the same as in Fig. \ref{fig:voltage-u-ol}.
}
\label{fig:voltage-u-la-sp}
\end{figure}

%
\begin{table}[htbp]
\begin{tabular}{|c|c|c|c|c|c|c|c|c|c|c|c|}
\hline  & Mn$^{2+}$ & Mn$^{3+}$ & Mn$^{4+}$ & Fe$^{2+}$ & Fe$^{3+}$ & Co$^{2+}$ & Co$^{3+}$ & Co$^{4+}$ & Ni$^{2+}$ & Ni$^{3+}$ & Ni$^{4+}$ \\
\hline Olivine & 3.92 & 5.09 &  & 3.71 & 4.90 & 5.05 & 6.34 &  & 5.26 & 6.93 &  \\
\hline Layered &  &  &  &  &  &  & 4.91 & 5.37 &  & 6.70 & 6.04 \\
\hline Spinel &  & 4.64 & 5.04 &  &  &  & 5.62 & 6.17 &  &  &  \\
\hline Monoxide \cite{Pickett-calcU1998} &3.6  &  &  & 4.6  &  & 5.0  &  &  & 5.1 &  &  \\
\hline
\end{tabular}
\caption{Calculated $U$ in eV.}
\label{tab:U}
\end{table}
Figures \ref{fig:voltage-u-ol} and \ref{fig:voltage-u-la-sp} show
respectively the average Li insertion voltage as function of $U$
in the olivine, and in the layered and spinel structure. The
horizontal short line indicates the experimentally measured
voltage. Three calculated points for each system are
 marked on the curve: the small open circles indicate respectively the voltage one would
obtain using the calculated $U$ for the most reduced and most oxidized TM-state in each structure
(e.g. Fe$^{2+}$ and Fe$^{3+}$ in LiFePO$_{4}$).
The large filled circle corresponds to the voltage for the averaged $U$. The results for each system are discussed in
more detail below.

\subsection{Olivine structures \olix \ (M=Mn, Fe, Co, Ni)}
According to neutron-diffraction experiments \cite{Santoro-AFM196x, Rousse-AFM2003} the magnetic
ordering of \oli\ is antiferromagnetic (AFM) within the approximately square lattice of metal ions
for each of the above four TM. FePO$_{4}$ is also found to have AFM magnetic ordering \cite{Rousse-AFM2003}.
The results in Fig. \ref{fig:voltage-u-ol} have been calculated with AFM spin configuration in both end members.
The calculated and experimental cell parameters, as well as the electronic occupation
of the TM ions are listed in Table \ref{tab:cell}.
\begin{table}[htbp]
\begin{tabular}{|p{57pt}|p{57pt}|p{53pt}|p{53pt}|p{53pt}|p{53pt}|p{78pt}|}
\hline& & a (\AA)& b (\AA)& c(\AA)& V(\AA$^{3}$)& TM ion config.
\\
\hline
LiMnPO$_4$ & GGA& 10.55& 6.13& 4.78& 309.13& $t_{2g}^3e_{g}^2$ \\\cline{2-7}  & GGA+$U$& 10.62& 6.17& 4.80& 314.52& $t_{2g}^3e_{g}^2$ \\\cline{2-7}  & Exp. \cite{Li-Mn_phosphate2002}& 10.44& 6.09& 4.75& 302.00&
\\
\hline
MnPO$_4$   & GGA& 9.92& 6.01& 4.93& 293.92& $t_{2g}^3e_{g}^1$ \\\cline{2-7}  & GGA+$U$& 9.98& 6.07& 4.96& 300.47& $t_{2g}^3e_{g}^1$ \\\cline{2-7}  & Exp.\cite{Li-Mn_phosphate2002} & 9.69& 5.93& 4.78& 274.67&
\\
\hline
LiFePO$_4$ & GGA& 10.39& 6.04& 4.75& 298.09& $t_{2g}^4e_{g}^2$ \\\cline{2-7}  & GGA+$U$& 10.42& 6.07& 4.76& 301.07& $t_{2g}^4e_{g}^2$ \\\cline{2-7}  & Exp. \cite{Goodenough1997} & 10.33& 6.01& 4.69& 291.39&  \\
\hline
FePO$_4$   & GGA& 9.99& 5.93& 4.90& 290.28& $t_{2g}^3e_{g}^2$ \\\cline{2-7}  & GGA+$U$& 9.99& 5.88& 4.87& 286.07& $t_{2g}^3e_{g}^2$ \\\cline{2-7}  & Exp. \cite{Goodenough1997} & 9.82& 5.79& 4.79& 272.36&  \\
\hline
LiCoPO$_4$ & GGA& 10.30& 5.93& 4.75& 290.13& $t_{2g}^5e_{g}^2$ \\\cline{2-7}  & GGA+$U$& 10.33& 5.97& 4.76& 293.55& $t_{2g}^5e_{g}^2$ \\\cline{2-7}  & Exp. \cite{Amine-Co_phosphate2000} & 10.20& 5.92& 4.70& 283.90&  \\
\hline
CoPO$_4$   & GGA& 9.71& 5.48& 4.59& 244.24& $t_{2g}^6$ \\\cline{2-7}  & GGA+$U$& 9.98& 5.78& 4.74& 273.42& $t_{2g}^4e_{g}^2$ \\\cline{2-7}  & Exp. \cite{Amine-Co_phosphate2000} & 10.09& 5.85& 4.72& 278.66&  \\
\hline
LiNiPO$_4$  & GGA& 10.09& 5.91& 4.74& 282.66& $t_{2g}^6e_{g}^2$ \\\cline{2-7}  & GGA+$U$& 10.12& 5.90& 4.73& 282.42& $t_{2g}^6e_{g}^2$ \\\cline{2-7}  & Exp. \cite{Garcia-LiNiPO4_cell2001} & 10.03& 5.85& 4.68& 274.49&  \\
\hline
NiPO$_4$& GGA& 9.66& 5.72& 4.71& 260.25& $t_{2g}^6e_{g}^1$ \\\cline{2-7}  & GGA+$U$& 9.92& 5.82& 4.84 & 279.43& $t_{2g}^6e_{g}^1$ \\
\hline
\end{tabular}
\caption{Cell parameters of the olivine structures in the lithiated and de-lithiated states,
as well as the corresponding electron configuration at the TM ions.}
\label{tab:cell}
\end{table}

\begin{description}
\item[Mn]
\
Both Mn$^{2+}$ and Mn$^{3+}$ are high-spin ions in GGA and GGA+$U$ calculations. Attempts to constrain them to lower spin states
lead to much higher energy. FM ordered magnetic structures are 10 - 30meV
higher in energy than the AFM ordered magnetic structure as $U$ is varied. A strong collective Jahn-Teller distortion is observed in
MnPO$_4$, where Mn$^{3+}$ is in the high-spin $t_{2g}^{3} e_{g}^{1}$ state, in GGA(+$U$).
The experimental voltage for the MnPO$_{4}$/LiMnPO$_{4}$ redox couple has been obtained from Ref.\ \onlinecite{Li-Mn_phosphate2002}.
The voltage predicted with GGA+$U$ (4.04 V at $U=(U_{{\rm Mn}^{2+}}+U_{{\rm Mn}^{3+}})/2$)
is within a few percent of the experimental voltage (4.1 V), and in sharp contrast to the large error made by GGA ($V_{\rm GGA}=2.98$ V).
\item[Fe]
\
Both Fe$^{2+}$ and Fe$^{3+}$ are high-spin in GGA(+$U$) calculations, and the AFM ordering is more stable than FM ordering.
Using $U_{{\rm Fe}^{2+}}$ and $U_{{\rm Fe}^{3+}}$  we calculated a voltage of 3.39 and 3.55 V respectively. The voltage calculated
with the average $U$= 4.30 eV is 3.47 V, which agrees very well with the experimentally measured value of 3.5 V \cite{Yamada-Fe_phosphate2001}. This is
a substantial improvement over the GGA predicted value of 2.97 V. Previously, the localization of electrons
induced by $U$ was also shown to qualitatively affect the phase behavior in this system \cite{Zhou-olivine2004}.
\item[Co] \ In LiCoPO$_{4}$ Co$^{2+}$ is stable in the high-spin
$t_{2g}^5 e_{g}^2$ state. In the delithiated CoPO$_{4}$, Co$^{3+}$
is stable as non spin-polarized with GGA, but more stable by
several eV with GGA+$U$ in the high spin $t_{2g}^4 e_{g}^2$
configuration at the calculated $U$ value of 6.34 eV. As shown in
Table \ref{tab:cell} the cell parameters of CoPO$_{4}$ calculated
with non spin-polarized Co$^{3+}$ in GGA is appreciably smaller
than experimental values, while GGA usually slightly overestimates
cell parameters. With GGA+$U$ and high-spin Co$^{3+}$ the
calculated parameters are close to experimental values. While
there is only limited electrochemical data on this material
\cite{Goodenough1997},
 the predicted voltage of 4.73 V at $U_{\rm average}$
is within a few \% of the result 4.8V established by Anime {\it et. al} \cite{Amine-Co_phosphate2000},
compared to the poor GGA prediction of 3.70 V. The high voltage of this material
makes it particularly attractive for high-energy density applications.
\item[Ni]
\
Though LiNiPO$_{4}$ has been synthesized, no Li can be removed from
it electrochemically \cite{Okada-Ni_phosphate2001}. Hence the voltage is probably
larger than 5V, the limit of most electrolyte systems.
At $x=1$ Ni$^{2+}$ is stable as high-spin $t_{2g}^6 e_{g}^2$.
At $x=0$ Ni$^{3+}$ occurs in the low spin state $t_{2g}^6 e_{g}^1$
for both GGA and GGA+U , but the high spin state $t_{2g}^5 e_{g}^2$ is less
unstable in GGA+U than in GGA. Note that low-spin Ni$^{3+}$ is a weak Jahn-Teller ion, and no appreciable collective
distortion is observed in our relaxed unit cell. With  $U_{\rm average}$, a voltage of 5.07 V is obtained, which is in
agreement with the fact that no Li can be removed from this material.
\end{description}

\subsection{Layered Li$_x$MO$_{2}$ (M=Co, Ni)}

For the layered and spinel structures AFM spin ordering on transition metal ions
is topologically frustrated, and their actual magnetic ground states are not always
clear in experiment. But as the energy associated with different magnetic orderings
is small, the simple FM ordering is used in the following calculations.
\begin{description}
\item[Co]
\
In LiCoO$_{2}$ Co$^{3+}$ is stable in the non spin-polarized state for the calculated
$U_{{\rm Co}^{3+}}=4.91$eV. At $x=0$, Co$^{4+}$ is almost degenerate in either
non spin-polarized or spin-polarized $t_{2g}^5$ in GGA, but more stable
with spin-polarization in GGA+U at the calculated $U_{{\rm Co}^{4+}}=5.37$eV.
While GGA+$U$ still improves the agreement of voltage with experiment \cite{Ohzuku-LA_Co1994}
over pure GGA, the error for this system is larger
than in the other systems we calculated.
This might be related to the fact that
the GGA result is already closer to experiment than for all other systems.
\item[Ni]
\ \
In LiNiO$_{2}$ Ni$^{3+}$ is most stable in the low-spin $t_{2g}^6 e_{g}^1$ state and is a weak Jahn-Teller ion.
With GGA a distorted unit cell is found with the short and the long Ni-O bond length being
1.92\AA\ and 2.13\AA, respectively, compared to experimental values of 1.91\AA\ and 2.14\AA\ \cite{Marianetti-JT-2001},
and a stabilization energy relative to an undistorted cell of only -2meV, within the range of numerical errors,
compared to -11 meV in Ref.\ \cite{Marianetti-JT-2001}.
With GGA+$U$ no appreciable distortion is observed. Experimentally there is no cooperative Jahn-Teller distortion in
LiNiO$_{2}$ though the Ni-O octahedra are locally Jahn-Teller distorted \cite{Rougier-Ni-JT1995},
suggesting a very small stabilization energy,
consistent with both GGA and GGA+$U$ results. At $x=0$,
Ni$^{4+}$ is stable as a non spin-polarized ion. The GGA+$U$
voltage value of 3.92V agrees well with the experimental average
voltage of 3.85V \cite{Delmas-LA_Ni1999}, and is substantially
better than the GGA result of 3.19V.
\end{description}

\subsection{Spinel Li$_{x}$M$_{2}$O$_{4}$ (M=Mn, Co)}
For the spinel  Li$_x$Mn$_2$O$_4$  there are two distinct plateaus in the voltage profile, between
$0<x<1$ and $1<x<2$, respectively. For $0 < x < 1$ Li enters tetrahedral sites, while
the reaction from LiMn$_{2}$O$_{4}$ to  Li$_{2}$Mn$_{2}$O$_{4}$ occurs through a two-phase process whereby the
 LiMn$_{2}$O$_{4}$ phase with only tetrahedral Li disappears at the expense of the  Li$_{2}$Mn$_{2}$O$_{4}$
phase with all Li octahedral.
 Calculations were done for $x=0$, 1 and 2 structures
to get separate average voltage values for the two processes. For M = Co the $0< x < 1$ reaction potential curve is difficult to
obtain accurately in experiments. Therefore only the average voltage for the $1 < x < 2$ reaction
is shown in fig. \ref{fig:voltage-u-la-sp}.
\begin{description}
\item[Mn]
\
Both Mn$^{4+}$ and Mn$^{3+}$ are high-spin. Mn$^{3+}$ is a strong Jahn-Teller active ion.
In GGA, the calculated Mn-O short and long bond lengths 1.94\AA\ and 2.40\AA\ agree with Ref. \onlinecite{Marianetti-JT-2001};
in GGA+$U$ they become 1.96\AA\ and 2.32\AA, respectively. Experimental values are 1.94\AA\ and 2.29\AA,
respectively \cite{Hoppe-SP_Mn-bond1975}, showing that the good structural prediction
of GGA is retained in GGA+$U$.  Coexistence of distinct
Mn$^{4+}$ and Mn$^{3+}$ is found in GGA+$U$ in the  LiM$_{2}$O$_{4}$ compound.
The GGA+$U_{\rm average}$ results (4.19V and 2.97V respectively, for the
first and second plateaus) is in excellent agreement with the experimentally measured values of
4.15V and 2.95V \cite{Ohzuku-LA_Co1994}.
\item[Co]
\
Like in the layered structure, Co$^{3+}$ in  Li$_{2}$Co$_{2}$O$_{4}$ is non spin-polarized, and at $x=0$ Co$^{4+}$ is more stable as spin polarized $t_{2g}^5$
in GGA+$U$. The GGA+$U$ voltage (3.56V at $U_{\rm average}$= 4.84eV ) agrees very well with experimental
data available for the Li$_{1}$Co$_{2}$O$_{4}$ to Li$_{2}$Co$_{2}$O$_{4}$ reaction (3.5V \cite{Choi-SP_Co2002}).
\end{description}
Note that in the $x=1$ structure of the spinel materials Li$_{x}$M$_{2}$O$_{4}$ we find distinct M$^{3+}$ and  M$^{4+}$
ions in GGA+U instead of ions of intermediate valence. The same phenomenon was observed in the intermediate structures
Li$_x$FePO$_4$ of the iron phosphate \cite{Zhou-olivine2004}. This is a direct consequence of the $E_{U}$ correction term
to the total energy in Eq. \ref{eq:ldapu} which penalizes the non-integral occupation of the $d$-orbitals. Such charge ordering is necessary
for correctly predicting the $0<x<1$ and $1<x<2$ average voltage values of Li$_x$Mn$_2$O$_4$ simultaneously,
as well as the $1<x<2$  voltage of Li$_x$Co$_2$O$_4$, and is not present in pure GGA unless localization is assisted
by a strong polaronic contribution such as the Jahn-Teller distortion around Mn$^{3+}$.

\section{Discussion \label{discussion}}

\begin{figure}[htbp]
\includegraphics[width=0.6 \linewidth, angle=-90]{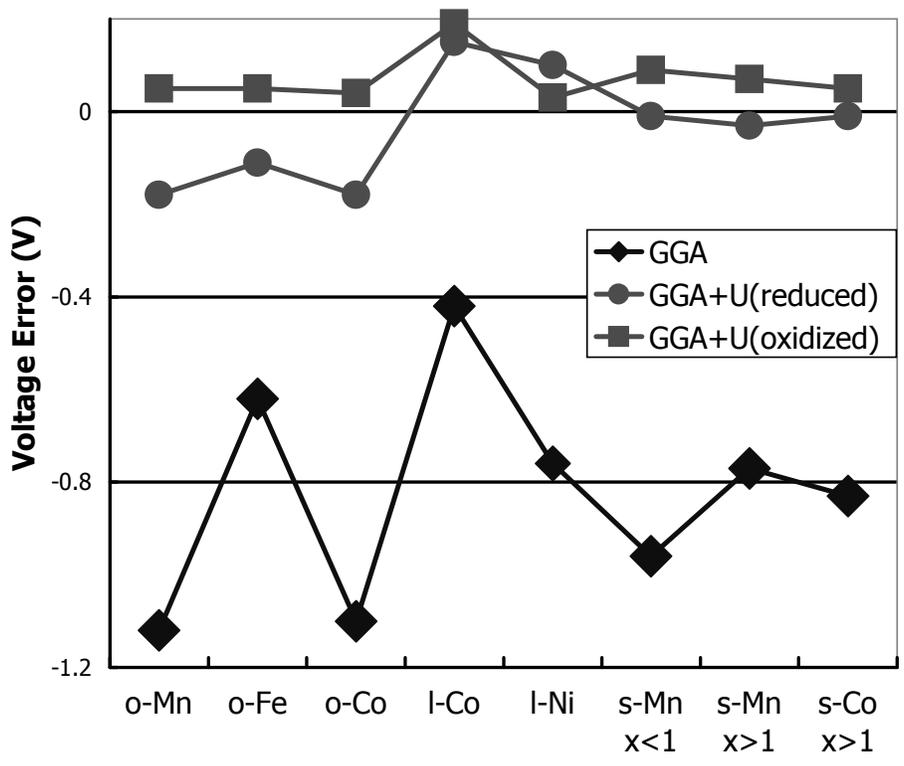}
\caption{
Difference between calculated and experimental voltage [5-9], for
GGA and GGA+U, at the   calculated
$U$ of the oxidized (delithiated) and reduced (lithiated) states, respectively (l=layered, s=spinel).
For the spinel structures two voltage values for the
$0< x < 1$ and $1< x < 2$ plateaus are calculated separately. Olivine LiNiPO$_4$ is not
shown here because the voltage is unknown.
}
\label{fig:errors}
\end{figure}

Introduction of Coulombic on-site correlations in GGA through the
GGA+$U$ clearly improves predicted lithiation potentials
considerably over the use of pure GGA (or LDA for that matter).
The errors of GGA+$U$ and pure GGA on all systems for which we
have experimental data are summarized in Fig. \ref{fig:errors}.
Pure GGA consistently underestimates the lithiation voltage, which
is a measure of the energy lowering when Li is transferred from Li
metal (the anodic reference) to a Li$^{+}$ ion and electron in the
TM oxide or phosphate. The contribution of the Li$^{+}$ ion to the
reaction energy is largely electrostatic, and one would expect
this effect to be well captured in GGA or LDA. Hence, the large
voltage error in LDA/GGA must arise from the electron transfer
from Li metal to the TM cation. Since the voltage is always
underestimated in LDA/GGA these approximations clearly penalize
the energy of the electron on the TM, thereby lowering the
reaction energy. It seems reasonable to attribute this to the poor
treatment of electronic correlations in LDA/GGA. In metallic
lithium the electron is affected by a small self-interaction in
LDA/GGA as its charge density is delocalized. On the TM ion, however, the
electron occupies a much more localized $d$-orbital and will
experience a much larger self-interaction.
The lack of cancellation between the self-interactions contributions to the
energy,
which are related to an improper description of the correlation
effects in LDA/GGA, leads to a
systematic error in the prediction of the redox potential. In the
direction in which the electron is transferred from a delocalized
to a localized state, the reaction energy is penalized (not
negative enough), making the potential too small. The use of
GGA+$U$ allows for a better description of the electronic correlation
and, by discouraging fractional occupations of the
Kohn-Sham orbitals, removes the spurious self-interaction
thus producing a much more accurate prediction of the redox
voltage.
While we demonstrate the
GGA/LDA problem and improvement obtained with DFT+$U$ on
Li-insertion materials, we believe that a more accurate description
of correlation effects within the DFT+$U$ scheme
is also necessary in the study of other redox processes
in which electrons are transferred between states of different kind
(e.g. catalysis of organic molecules on TM surfaces).
In fact, as explained in Ref. \onlinecite{Cococioni-thesis2002},
a better description of the electronic correlation
(which enforces the independence of the single electron
energy eigenvalues of the partially occupied states
on their occupation, thus leading to the elimination of the spurious
self-interaction) is needed to
reproduce the physical difference among the ionization
potential and the electronic affinity (or the band gap in crystalline
solids) which plays a very important role in the energetics
of processes involving electron transfer.

In our calculations high-spin TM ions are always energetically
favored by GGA+$U$ over low-spin or non spin-polarized states. In
CoPO$_4$ the non spin-polarized Co$^{3+}$ in GGA leads to cell
parameters inconsistent with experiment. In GGA+$U$ Co$^{3+}$
becomes high spin, improving agreement with experiment.
For the other systems the GGA and GGA+$U$
cell parameters are rather close, though GGA+$U$ seems to lead to volumes
that are slightly too high. Jahn-Teller distortions predicted by GGA are also reproduced in GGA+$U$ for Mn$^{3+}$.

In summary, we have shown that the under-estimation of the lithium intercalation
voltage in LDA/GGA can be corrected by using GGA+$U$ with
a self-consistently calculated  parameters $U$,
without sacrificing properties that are already accurately predicted by GGA
(e.g. Jahn-Teller effect, cell parameters, magnetic ordering).
Voltages for most systems are predicted within a few \% of experimental values.

We believe that DFT+$U$ will significantly improve the accuracy
of voltage prediction for candidate materials can be predicted, and
therefore enhance the capability of screening new materials for their
ability to be good cathodes.
\vskip 1cm

\begin{acknowledgments}
The authors thank Thomas Maxisch for helpful discussions.
This work was supported by the Department of Energy under Contract
number DE-FG02-96ER45571 and by the MRSEC program of the National
Science Foundation under contract number DMR-0213282.
\end{acknowledgments}

\end{document}